\documentclass[a4paper,fleqn,usenatbib]{mnras}

\usepackage{newtxtext,newtxmath}

\usepackage[T1]{fontenc}
\usepackage{ae,aecompl}

\usepackage{graphicx}	
\usepackage{amsmath}	
\usepackage{amssymb}	

\usepackage[textsize=footnotesize]{todonotes}
\usepackage{hyperref}

\newcommand{\chandra}{{\it Chandra}}

\newcommand{\hst}{{\it HST}}

\newcommand{\lsun}{${\rm L_{\sun}}$}
\newcommand{\msun}{${\rm M_{\sun}}$}

\newcommand{\kms}{${\rm km~s^{-1}}$}

\newcommand{\aco}{CO(1--0)}
\newcommand{\bco}{CO(2--1)}
\newcommand{\cco}{CO(3--2)}
\newcommand{\dco}{CO(4--3)}
\newcommand{\eco}{CO(5--4)}

\title[A CO survey of blended submillimetre sources]{An ALMA survey of CO in submillimetre galaxies: companions, triggering, and the  environment in blended sources}

\author[J.\ L.\ Wardlow et al.]{Julie L.\ Wardlow$^{1}$\thanks{E-mail: julie.wardlow@durham.ac.uk},
J.\ M.\ Simpson$^{2}$,
Ian Smail$^{1}$,
A.\ M.\ Swinbank$^{1}$,
A.\ W.\ Blain$^{3}$,
\newauthor
W.\ N.\ Brandt$^{4,5,6}$,
S.\ C.\ Chapman$^{7}$,
Chian-Chou Chen$^{8}$,
E.\ A.\ Cooke$^{1}$,
\newauthor
H.\ Dannerbauer$^{9,10}$,
B.\ Gullberg$^{1}$,
J.\ A.\ Hodge$^{11}$,
R.\ J.\ Ivison$^{8,12}$,
K.\ K.\ Knudsen$^{13}$,
\newauthor
Douglas Scott$^{14}$,
A.\ P.\ Thomson$^{15,1}$, 
A.\ Wei\ss$^{16}$, and
P.\ P.\ van der Werf$^{11}$
\\
$^{1}$ Centre for Extragalactic Astronomy, Department of Physics, Durham University, South Road, Durham, DH1 3LE, UK\\
$^{2}$ Academia Sinica Institute of Astronomy and Astrophysics, No. 1, Sec. 4, Roosevelt Rd., Taipei 10617, Taiwan\\
$^{3}$ University of Leicester, Physics and Astronomy, University Road, Leicester, LE1 7RH, UK\\
$^{4}$ Department of Astronomy and Astrophysics, 525 Davey Lab, The Pennsylvania State University, University Park, PA 16802, USA \\
$^{5}$ Institute for Gravitation and the Cosmos, The Pennsylvania State University, University Park, PA 16802, USA \\
$^{6}$ Department of Physics, 104 Davey Laboratory, The Pennsylvania State University, University Park, PA 16802, USA \\
$^{7}$ Department of Physics and Atmospheric Science, Dalhousie University, Halifax, NS B3H 3J5 Canada\\
$^{8}$ European Southern Observatory, Karl Schwarzschild Strasse 2, Garching, Germany \\
$^{9}$ Instituto de Astrof{\'i}sica de Canarias (IAC), E-38205 La Laguna, Tenerife, Spain\\
$^{10}$ Universidad de La Laguna, Dpto. Astrof'sica, E-38206 La Laguna, Tenerife, Spain \\
$^{11}$ Leiden Observatory, Leiden University, P.O. Box 9513, NL- 2300 RA Leiden, The Netherlands\\
$^{12}$ Institute for Astronomy, University of Edinburgh, Royal Observatory, Blackford Hill, Edinburgh EH9 3HJ, UK \\
$^{13}$ Department of Space, Earth and Environment, Chalmers University of Technology, Onsala Space Observatory, SE-439 92 Onsala, Sweden \\
$^{14}$ Department of Physics and Astronomy, 6224 Agricultural Road, University of British Columbia, Vancouver V6T 1Z1, Canada\\
$^{15}$ Jodrell Bank Centre for Astrophysics, The University of Manchester, Oxford Road, Manchester, M13 9PL, UK\\
$^{16}$ Max-Planck-Institut f\"ur Radioastronomie, Auf dem H\"ugel 69, D-53121 Bonn, Germany
}

\date{Accepted XXX. Received YYY; in original form ZZZ}

\pubyear{2018}

\begin{document}
\label{firstpage}
\pagerange{\pageref{firstpage}--\pageref{lastpage}}
\maketitle

\begin{abstract}
We present ALMA observations of the mid-$J$ $^{12}$CO emission from six single-dish selected 870-\micron\ sources in the Extended {\it Chandra} Deep Field-South (ECDFS) and UKIDSS Ultra-Deep Survey (UDS) fields. These six single-dish submillimetre sources were selected based on previous ALMA continuum observations, which showed that each comprised a blend of emission from two or more individual submillimetre galaxies (SMGs), separated on 5--10\arcsec\ scales. The six single-dish submillimetre sources targeted correspond to a total of 14 individual SMGs, of which seven have previously-measured robust optical/near-infrared spectroscopic redshifts, which were used to tune our ALMA observations. 
We detect \cco\ or \dco\ at $z=2.3$--3.7 in seven of the 14 SMGs, and in addition serendipitously detect line emission from three gas-rich companion galaxies, as well as identify four new 3.3-mm selected continuum sources in the six fields.
Joint analysis of our CO spectroscopy and existing data suggests that $64(\pm18)\%$ of the SMGs in blended submillimetre sources are unlikely to be physically associated. 
However, three of the SMG fields (50\%) contain new, serendipitously-detected CO-emitting (but submillimetre-faint) sources at similar redshifts to the 870-\micron\ selected SMGs we targeted. These data suggest that the SMGs inhabit overdense regions, but that these are not sufficiently overdense on $\sim100$\,kpc scales to influence the source blending given the short lifetimes of SMGs. 
We find that $21\pm12\%$ of SMGs have spatially-distinct and kinematically-close companion galaxies ($\sim8$--150\,kpc and $\lesssim300$\,\kms), which may have enhanced their star-formation via gravitational interactions.
\end{abstract}

\begin{keywords}
galaxies: evolution -- submillimetre: galaxies -- galaxies: ISM 
\end{keywords}



\section{Introduction}
\label{sec:intro}

Luminous submillimetre-selected galaxies, with flux densities above 1--2\,mJy in (sub)millimetre observations \citep[SMGs; e.g.][]{Smail97, Barger98, Hughes98, Coppin06, Scott08, Weiss09, Geach17}, typically have  infrared luminosities of $L_{\rm IR}\sim10^{12-13}$\,\lsun, corresponding to star-formation rates of hundreds to thousands of solar masses per year \citep[e.g.][]{Chapman05, Pope06, Wardlow11, Swinbank14}. They have inferred stellar masses of $\sim10^{10.8-11.1}$\,\msun\ \citep[e.g.][]{Hainline11, Wardlow11, Michalowski12, Simpson14}, are the most actively star-forming galaxies at $z\sim2$, where the redshift distribution of 850-\micron\ selected SMGs peaks \citep[e.g.][]{Chapman05, Pope05, Wardlow11, Simpson14, Zavala14}, and they contribute around 20\% of the cosmic star-formation rate density at this epoch \citep[for flux density at 870\,\micron, $S_{870}>1$\,mJy; e.g.][]{Barger12, Swinbank14}. Clustering analyses have shown that SMGs typically reside in overdense regions (halo masses $\sim10^{13}$\,\msun\ at $z\gtrsim2$), and likely evolve into $\gtrsim L^*$ galaxies with stellar masses of $\sim10^{11-11.3}$\,\msun\ in local clusters \citep{Hickox12, Simpson14, Hodge16, Wilkinson17}. 
Luminous SMGs are thus potentially a significant component of the stellar mass growth of the Universe and a key phase in the formation of local early type-galaxies.

Despite nearly 20\,years of study, the physical process responsible for triggering the activity in SMGs is still a subject of intense debate. Morphologically the dusty (i.e.\ star-forming) regions in SMGs, as traced by high-resolution submillimetre interferometry, are typically small (a few kpc) discs \citep[e.g.][]{Younger08, Simpson15a, Hodge16}, but the rest-frame optical emission, from \hst\ data, displays a wide range of morphologies including discs, apparent spheroids, irregular systems, possible interactions and unresolved galaxies \citep[e.g.][]{Swinbank10b, Wiklind14, Chen15}. Dynamical studies also provide a mixed picture with rotating discs, disturbed systems and possible mergers all observed \citep[e.g.][]{Tacconi08, Engel10, Riechers11b, Hodge12, Ivison13}.
Theoretical predictions from simulations differ significantly, with many numerical and some hydrodynamical simulations \citep[e.g.][]{Hayward12, Hayward13a, Narayanan10} identifying merger activity as the predominant trigger for bright SMGs ($\sim90\%$ for $S_{870}>5$\,mJy SMGs in \citealt{Hayward13a}), but with a submillimetre flux dependance such that secular processes drive the star formation in less luminous systems. By contrast, the most recent version of the semi-analytic model of galaxy formation, {\sc galform} \citep{Cowley15, Lacey16}, and some hydrodynamical simulations \citep[e.g.][]{Dave10, Hayward11, Narayanan15}, predict that bright SMGs typically represent isolated, gas-rich, Toomre-unstable discs undergoing intense secular bursts.

Recent Atacama Large Millimeter/submillimeter Array (ALMA) continuum follow-up studies of the submillimetre sources identified in wide-field single-dish surveys  have demonstrated the importance of blending in the coarse single-dish beams, with 35--60\% of bright single-dish submillimetre sources found to comprise two or more individual SMGs, when observed at arcsecond resolution and to sub-mJy rms depths \citep[e.g.][]{Karim13, Hodge13, Simpson15b, Miettinen15}, confirming earlier suggestions using radio emission as a far-infrared proxy \citep[e.g.][]{Ivison07}. 
Blending is compounded by the negative K-correction at submillimetre wavelengths, which means that fixed luminosity galaxies across a broad range of redshifts ($z\sim1$--8) are similarly easy to detect, and that the blending of multiple SMGs into a single-dish submillimetre source can arise from galaxies at substantially different redshifts.
It is also important to note that the detected blending rate is dependent on both the rms and synthesised beam of the high-resolution observations, as well as the brightness of the single-dish submillimetre sources observed (e.g.\ \citealt{Smolcic12, Hill17}; Stach et al.\ 2018, in prep.).

High-resolution hydrodynamical simulations have shown that both early-stage (i.e.\ wide-separation) mergers and companion galaxies in the local SMG environment can contribute to the blending of submillimetre sources, although these simulations have not addressed whether such situations are a dominant contributor to the blended submillimetre source population \citep[e.g.][]{Narayanan10, Narayanan15}. 
However, a range of large-scale statistical models (with differing explanations for SMG triggering) predict that blended submillimetre sources are mostly comprised of physically-unassociated, chance, line-of-sight alignments of SMGs \citep{Hayward13b, Cowley15, MunozArancibia15}. 
This expectation is in tension with the observed space density of blended, multiple component submillimetre sources in ALMA data, which is a factor of $\sim80$ higher than can be explained by chance alone, based on blank-field submillimetre number counts \citep{Simpson15b}, indicating that the blended SMGs may often be  associated, and perhaps that an interaction between the components triggered their starbursts. However, this assumption has not yet been robustly tested with redshift information because the typically-available photometric redshifts do not have the required precision.
\citet{Hayward18} used optical and near-infrared spectroscopy to examine the nature of multiplicity in  ten submillimetre sources, but were dogged by incompleteness in the redshift information, and were only able to determine that $\sim50\%$ of their sample contained at least one physically-unassociated SMG. 
The challenge of optical/near-infrared spectroscopy for blended SMGs is due to their high-redshifts, significant dust absorption and the necessary close spacing of slits \citep[see e.g.][]{Danielson17}. 
 Instead, another route to testing whether blended SMGs are physically associated is via spatially-resolved millimetre spectroscopy to detect the molecular gas emission, where the multiple components can be observed simultaneously and dust absorption has a negligible effect on the ability to determine redshifts. 

In this paper we investigate whether the blended multiple SMG components of single-dish submillimetre sources are physically associated via an ALMA $^{12}$CO survey in Band 3, and VLT/XSHOOTER spectroscopic data. 
We also use the ALMA data to identify gaseous companions to the SMGs and determine the fraction that are triggered by interactions with dust/gas-rich companions. Detailed analyses of the individual galaxies will be presented in a forthcoming paper (Wardlow et al.\ in prep.).

This paper is organised as follows: in Section~\ref{sec:obs} we describe the sample selection, observations, and data
reduction. Section~\ref{sec:results} includes our main results and Section~\ref{sec:analysis} contains the analysis and discussion. Our conclusions are presented in
Section~\ref{sec:conc}.  Throughout this paper we use $\Lambda$CDM
cosmology with $\Omega_{\rm M}=0.286$, $\Omega_{\Lambda}=0.714$ and
$H_{0}=69.6\,{\rm km\,s^{-1}\,Mpc^ {-1}}$ \citep{Wright06, Bennett14}.

\begin{table*}
 \caption{Basic information and measurements of the SMG targets and serendipitously-identified systems}
 \label{tab:sample}
 \setlength{\tabcolsep}{3pt} 
  \begin{tabular}{llcccccccc}
  \hline\hline
	Source & Name$^a$ & Position$^a$ & $S_{870}$$^a$ & $z_{\rm opt}$$^b$ & $S_{3.3\rm mm}$$^c$ & $I_{\rm CO}$$^d$ & FWHM$^e$ & $z_{\rm \sc co}$$^f$  & $L^{\prime}_{\rm CO}$$^g$ \\
		&            & (J2000) & (mJy)         &              &   ($\mu$Jy)  &   (Jy~\kms)  &   (\kms)  &  & ($10^{10}$~K~\kms~pc$^2$) \\
  \hline  \hline
LESS\,41 & ALESS\,41.1 & 03$^{\rm h}$31$^{\rm m}$10\fs07 $-$27\degr52\arcmin36\farcs7 &$4.9\pm0.6$ & 2.5460& $66\pm3$ & $0.85\pm0.03$& $700\pm30$ & $2.5470\pm0.0001$ & $2.9\pm0.1$ \\
         & ALESS\,41.3 & 03$^{\rm h}$31$^{\rm m}$10\fs30 $-$27\degr52\arcmin40\farcs8 &$2.7\pm0.8$ & \ldots& $<12$ & ${\it<0.11}$ & \ldots  & \ldots &  ${\it<0.38}$ \\
         & ALESS\,41.C & 03$^{\rm h}$31$^{\rm m}$09\fs81 $-$27\degr52\arcmin25\farcs4 & $3.2\pm1.6$ &\ldots & $198\pm6$ & \ldots & \ldots  &\ldots &\ldots\\
    \hline
LESS\,49 & ALESS\,49.1 & 03$^{\rm h}$31$^{\rm m}$24\fs72 $-$27\degr50\arcmin47\farcs1 &$6.0\pm0.6$ & 2.9417& $37\pm5$ & $0.88\pm0.03$& $590\pm60$ & $2.9451\pm0.0003$& $3.9\pm0.1$ \\
         & ALESS\,49.2 & 03$^{\rm h}$31$^{\rm m}$24\fs47 $-$27\degr50\arcmin38\farcs1 &$1.8\pm0.4$ & \ldots& $28\pm6$ & ${\it<0.07}$ & \ldots &\ldots & ${\it <0.32}$ \\  
         & ALESS\,49.C & 03$^{\rm h}$31$^{\rm m}$24\fs58 $-$27\degr50\arcmin43\farcs4 & $<1.3$ & \ldots & $43\pm5$ & \ldots&\ldots & \ldots& \ldots \\
         & ALESS\,49.L & 03$^{\rm h}$31$^{\rm m}$24\fs72 $-$27\degr50\arcmin43\farcs7 & $<1.3$ & \ldots& $<15$ & $0.16\pm0.02$& $550\pm60$& $2.9300\pm0.0003$ & $0.68\pm0.07$\\
    \hline
LESS\,71 & ALESS\,71.1 & 03$^{\rm h}$33$^{\rm m}$05\fs65 $-$27\degr33\arcmin28\farcs2 &$2.9\pm0.6$ & 3.6967& $<33$ & $1.25\pm0.07$ & $630\pm50$& $3.7089\pm0.0002$ & $4.5\pm0.2$\\
         & ALESS\,71.3 & 03$^{\rm h}$33$^{\rm m}$06\fs14 $-$27\degr33\arcmin23\farcs1 &$1.4\pm0.4$ & \ldots& $<33$ & ${\it<0.25}$ & \ldots & \ldots & ${\it <0.92}$ \\
    \hline
LESS\,75 & ALESS\,75.1 & 03$^{\rm h}$31$^{\rm m}$27\fs19 $-$27\degr55\arcmin51\farcs3 &$3.2\pm0.4$ & 2.5450& $28\pm2$& $0.99\pm0.02$& $530\pm10$ & $2.5521\pm0.0001$ & $3.43\pm0.09$\\
         & ALESS\,75.2$^h$ & 03$^{\rm h}$31$^{\rm m}$27\fs67 $-$27\degr55\arcmin59\farcs2 &$5.0\pm1.2$ & 2.2944& $<7$& \ldots& \ldots &\ldots & \ldots \\
         & ALESS\,75.4 & 03$^{\rm h}$31$^{\rm m}$26\fs57 $-$27\degr55\arcmin55\farcs7 &$1.3\pm0.4$ & \ldots& $<7$& ${\it<0.10}$ & \ldots  & \ldots & ${\it <0.33}$\\
         & ALESS\,75.C & 03$^{\rm h}$31$^{\rm m}$26\fs65 $-$27\degr56\arcmin01\farcs1 & $<1.0$ & $4.00^{+0.07}_{-0.08}$ & $58\pm3$& \ldots& \ldots &\ldots & \ldots\\
    \hline
LESS\,87 & ALESS\,87.1 & 03$^{\rm h}$32$^{\rm m}$50\fs88 $-$27\degr31\arcmin41\farcs5 &$1.3\pm0.4$& 2.3086& $46\pm7$& $0.40\pm0.03$ & $680\pm60$& $2.3136\pm0.0002$ & $1.17\pm0.08$ \\
         & ALESS\,87.3 & 03$^{\rm h}$32$^{\rm m}$51\fs27 $-$27\degr31\arcmin50\farcs7 &$2.4\pm0.6$ & \ldots & $<44$& ${\it<0.11}$ & \ldots &\ldots &${\it <0.33}$\\
         & ALESS\,87.C & 03$^{\rm h}$32$^{\rm m}$50\fs65 $-$27\degr31\arcmin34\farcs9 &  $<1.8$ & \ldots& $79\pm7$& \ldots &\ldots &\ldots &\ldots\\
         & ALESS\,87.L & 03$^{\rm h}$32$^{\rm m}$52\fs42 $-$27\degr31\arcmin49\farcs1 &  \ldots & \ldots& $35\pm11$& $0.31\pm0.03$& $450\pm70$ & $2.3141\pm0.0002$ &$0.92\pm0.07$ \\
    \hline
UDS\,306$^i$ & UDS\,306.0 & 02$^{\rm h}$17$^{\rm m}$17\fs07 $-$05\degr33\arcmin26\farcs6 &$8.3\pm0.5$ & 2.603& $98\pm9$& $2.18\pm0.06$ & $560\pm20$& $2.5991\pm0.0003$ & $7.8\pm0.3$\\
         & UDS\,306.1 & 02$^{\rm h}$17$^{\rm m}$17\fs16 $-$05\degr33\arcmin32\farcs5 &$2.4\pm0.4$ & \ldots & $83\pm17$& $0.60\pm0.02$ & $400\pm20$& $2.6136\pm0.0001$& $2.17\pm0.08$\\
         & UDS\,306.2$^i$ & 02$^{\rm h}$17$^{\rm m}$16\fs81 $-$05\degr33\arcmin31\farcs8 &$2.3\pm0.9$ & \ldots & $<51$& ${\it<0.14}$ & \ldots &\ldots &${\it <0.49}$ \\
         & UDS\,306.L & 02$^{\rm h}$17$^{\rm m}$17\fs10 $-$05\degr33\arcmin31\farcs5 & $<0.7$& $ \ldots $ & $<53$& $0.59\pm0.03$ & $370\pm20$ & $2.6132\pm0.0001$& $2.1\pm0.1$\\
  \hline  \hline
\end{tabular}
 \\{\bf Notes -- } 
 $^a$ Names, positions and 870-\micron\ continuum fluxes ($S_{870}$) are from \citet{Hodge13} and \citet{Simpson15b} for  870-\micron\ selected SMGs. Exceptions are sources with suffixes `C' and `L', which denote serendipitously identified 3.3\,mm continuum and line emitters, respectively (Sections~\ref{sec:serendip_cont} and \ref{sec:serendip_line}), for which the $3\sigma$ limits or newly-measured 870-\micron\ flux densities are reported where possible. 
 $^b$ Redshifts from optical or near-infrared spectroscopy \citep[][Chapman et al.\ in prep.; Section~\ref{sec:serendip_cont}]{Danielson17}.  
 $^c$ 3.3-mm continuum flux densities; $3\sigma$ limits are presented for sources not detected above this level. 
$^d$ Integrated line flux: $S_{\rm CO}\Delta v$. For SMGs without detected lines we present the detection limits (in italics) for a spatially-unresolved 500\,\kms\ line {\it within the observed frequency range}.
$^e$ For lines that are fit with a double-Gaussian profile (Figure~\ref{fig:overview}), we provide here the equivalent FHWM, calculated from the width containing 68\% of the line flux. 
$^f$ Source redshift, assuming that the detected line is the same CO line as for the primary SMG target, i.e.\ \dco\ for ALESS71.1 and \cco\ for all other sources. 
$^g$ Observed CO line luminosity (i.e.\ \dco\ for LESS71 and \cco\ for all other fields). For CO-undetected 870-\micron\ selected SMGs the limits (in italics) are correct {\it only if the SMG is at the same redshift as the spectroscopically confirmed SMGs in the field}.  
$^h$ LESS\,75.2 is a less-reliable {\sc Supplementary} source in the \citet{Hodge13} catalogue, and our ALMA tunings do not cover the frequency of the expected CO line for the optical redshift.
$^i$ UDS\,306 \citep{Simpson15b} is named UDS\,17 in the final version of the S2CLS catalogue \citet{Geach17} and subsequent ALMA catalogue \citep{Simpson15b}. 
 \end{table*}

\section{Observations and data reduction}
\label{sec:obs}

\subsection{ALMA data}
\label{sec:almaobs}

Our ALMA targets were selected from two blank-field, single-dish, 870-\micron\ surveys: the LABOCA ECDFS Submillimetre Survey \citep[LESS;][]{Weiss09} and the observations of the Ultra-Deep Survey (UDS) field from the SCUBA-2 Cosmology Legacy Survey \citep[S2CLS;][]{Geach17}. Both LESS and S2CLS-UDS have ALMA followup observations of the single-dish identified SMGs (ALESS: \citealt{Hodge13}; AS2UDS: \citealt{Simpson15b} and Stach et al.\ 2018, in prep.), as well as significant optical/near-infrared spectroscopic observations (Section~\ref{sec:xshooterobs}; \citealt{Danielson17}; Chapman et al.\ in prep.). 

We selected single-dish 870-\micron\ sources that meet the following criteria: (a) have multiple robust components in the ALMA continuum follow-up \citep[][Stach et al.\ 2018, in prep.]{Hodge13, Simpson15b}; (b) have a robust (i.e.\ multiple-line) optical/near-infrared spectroscopic redshift for (at least) one of the ALMA galaxies \citep{Danielson17}; (c) without spectroscopic redshift(s) for the other ALMA galaxy(ies) that are blended in the single-dish data;
and (d) have an ALMA-accessible CO line at the spectroscopic redshift.
Based on these criteria our ALMA targets comprised five LESS and one S2CLS-UDS single-dish sources, which between them  comprise 14 870-\micron\ ALMA-identified SMGs (Table~\ref{tab:sample}).

Ten of our 14 target SMGs (71\%) have detectable optical, near-infrared and/or mid-infrared counterparts, which is consistent with the 80\% of the parent ALESS sample with multiwavelength counterparts \citep{Simpson14}. \citet{Simpson14} used stacking to show that the ALESS sources that are not individually-detected in the multiwavelength data have faint optical/infrared counterparts, and that the majority are likely to be real sources in the 870-\micron\ ALMA data. We therefore expect that the majority of our targeted SMGs are real galaxies, although our conclusions would not change if one of the undetected SMGs are actually spurious. 
Also, note that requirement (c) does not significantly bias our sample, since, due to the difficulty in obtaining spectroscopic optical/near-infrared redshifts for SMGs, only one blended submillimetre source in our parent sample has existing spectroscopic redshifts for all component SMGs. We include this source in the analyses in Section ~\ref{sec:analysis}. 

The primary goal of the ALMA observations is to determine whether the secondary (and tertiary) SMG components are at the same redshift as the spectroscopically confirmed-target (i.e.\ whether or not the multiple components are physically associated) via the presence or absence of $^{12}$CO emission at the redshift of the primary component. 
The data are from ALMA Project 2016.1.00754 and were taken in ALMA Band 3 (at $\sim3.3$~mm or $\sim90$~GHz) and tuned such that one of the $\sim2$\,GHz sidebands included the expected \cco\ line (LESS\,41, LESS\,49, LESS\,75, LESS\,87 and UDS\,306 fields) or \dco\ line (LESS\,71 field only)  at the redshift of the spectroscopically-confirmed component. The observations were taken during Cycle 4 on 2016 November 12, 16 and 20 and have angular resolutions of 0.8--1.1\arcsec. In these configurations the Maximum Recoverable Scale is $\sim11\arcsec$ so we expect to detect all of the flux from these high-redshift galaxies (typical far-infrared and gas sizes of SMGs are $\sim0.2$--1.2\arcsec; \citealt{Tacconi06, Engel10, Ivison10, Ikarashi15, Simpson15a, Hodge16}).  
Several of the SMGs have marginally resolved $^{12}$CO and continuum emission in our data, which will be discussed further in Wardlow et al.\ (in prep.). The FWHM of the primary beam is $\sim55\arcsec$ (i.e.\ our Band 3 maps cover $\sim10\times$ the area of the ALMA 870-\micron\ maps) and the pointings were centred between the individual SMG components. 
Integration times ranged from $\sim800$ to $\sim6400$~s and are sufficient to detect $^{12}$CO in the secondary (and tertiary) components at $>5\sigma$ if they are at the same redshifts as the primary components and follow the typical far-infrared to CO luminosity relation (Section~\ref{sec:blends}).
The bandpass and phase calibrators  were J0334$-$4008 and J0317$-$2803 or J0334$-$4008 and J0343$-$2530 for the LESS fields, and J0238$+$1636 and J0217$-$0820 for the UDS field, and these were also used to calibrate the fluxes.

For data reduction and analysis we use the Common Astronomy Software Applications ({\sc casa}) version 4.7.0, using the ALMA-provided pipeline scripts to reduce the data. The data were imaged in stages using the {\sc clean} task. We initially used natural weighting (i.e.\ Briggs weighting with `${\rm Robust=2}$') with no cleaning to generate a `dirty' cube, which was used to identify any serendipitous emission line sources (Section~\ref{sec:serendip_line})\footnote{The strength of the SMG emission from UDS306.0, combined with the poorer {\it uv}-coverage in this field, led to significant sidelobes, so this cube was {\sc clean}ed prior to the search for serendipitous sources.} and CO from the SMGs (Section~\ref{sec:findsmglines}). We next subtracted the channels containing emission lines from the data and collapsed the cube to generate a naturally-weighted, dirty continuum map, which was used to identify additional continuum sources (Section~\ref{sec:serendip_cont}).

To generate line-free continuum maps we excluded the channels containing emission lines and used {\sc clean} to image the remaining data, manually masking sources and {\sc clean}ing the data to approximately rms level. For the purposes of this work we created naturally-weighted maps, which have 3.3-mm continuum noise of 2--$11~\mu{\rm Jy~beam^{-1}}$, with synthesized beams typically 0.8--1.1\arcsec\ (Figure~\ref{fig:overview}).
The ALMA 3.3~mm continuum flux densities (or limits) for the 870-\micron\ selected SMGs and serendipitously-identified sources were measured from the primary beam corrected, naturally-weighted maps using the {\sc casa} task {\sc imfit} (Table~\ref{tab:sample}). 

Finally, we produced continuum-subtracted cubes for analysis of the CO emission. Using {\sc uvcontsub}, in the {\it uv}-plane, we fitted the continuum and subtracted it from sidebands containing SMG line emission or the serendipitously-detected emission line sources (Section~\ref{sec:serendip_line}). The sources were masked and these continuum-subtracted data were manually {\sc clean}ed, to generate naturally-weighted  maps for analysis.

\subsection{XSHOOTER spectroscopy}
\label{sec:xshooterobs}

Spectroscopic observations of eleven ALMA SMGs in the UDS field from  S2CLS \citep{Geach17} were made with the
XSHOOTER spectrograph \citep{Vernet11} on the ESO VLT/UT2 between 2014 November 20 and 2015
January 19 as part of programme 094.A-0811.  These targets were
selected from ALMA maps of (bright) SCUBA-2 sources that, from the pilot survey, contain two
or more ALMA-identified SMGs, each with fluxes $>$2\,mJy from \citet{Simpson15b}. 
In total we placed slits on ten ALMA-identified SMGs, associated with five different SCUBA-2 sources. 

XSHOOTER was used in its cross-dispersed mode, which results in
simultaneous wavelength coverage of $\sim$\,300--2500\,nm at a
resolution $R=\lambda/\Delta\lambda=5000$--10000.
Observations were made in dark time in clear atmospheric conditions.
We set the slit width to approximately match the seeing (0.9\arcsec\ in the
visible and 0.7\arcsec\ in the near-infrared).  In the cross-dispersed
mode, the slit length is 11\arcsec\ and so each target was first centred
on the slit using an offset from a nearby star (rotated to the
parallactic angle at the mid-point of the observation).  For sky
subtraction, we nodded the slit $\pm2\arcsec$ along the source.  Each
target was observed for 7200\,s, split into 900\,s exposures.

Data reduction was carried out using the {\sc EsoRex}\footnote{\url{www.eso.org/sci/software/cpl/esorex.html}} pipeline, which
extracts, wavelength calibrates and sky-subtracts each observation;
these were then coadded to create a final two-dimensional stacked
spectrum.  One dimensional spectra were extracted, and both the
one- and two-dimensional spectra were inspected to search for emission
lines.  Spectra for other sources that serendipitously fell on the slits were also extracted where
continuum and/or emission lines were detected. 

Of the ten SMGs observed with XSHOOTER, six exhibit no strong emission lines, and thus redshifts cannot be determined for these sources (UDS\,48.0, UDS\,48.1, UDS\,269.0, UDS\,269.1, UDS\,286.2, and UDS\,298.1; all source names from \citealt{Simpson15b}). Three of the targets in UDS\,286 (UDS\,286.0, UDS\,286.2, and UDS\,286.3) contain emission from the same foreground galaxy at $z=0.44$, which is offset from the SMGs, but falls in the slits due to small separation of the three SMGs and the position angles of the observations. No line emission from these SMGs is observed and therefore their redshifts cannot be determined from the XSHOOTER data. Finally, the observation of UDS\,306.2 fails to detect any emission from the SMG, but a source at $z=2.605$ , approximately 3\arcsec\ from the SMG (with RA, Dec of 02$^{\rm h}$17$^{\rm m}$16\fs65, $-$05\degr33\arcmin29\farcs9, respectively), is serendipitously detected by XSHOOTER. Note that this redshift is close to that for UDS\,306.0, UDS\,306.1 and UDS\,306.L from  ALMA  (Table~\ref{tab:sample}), and indicates that the structure traced by the two 870-\micron\ selected SMGs and one gas-rich system (Section~\ref{sec:environ}) contains at least one additional gas/dust-poor galaxy.

\section{Results}
\label{sec:results}

\begin{figure*}
\includegraphics[width=\textwidth]{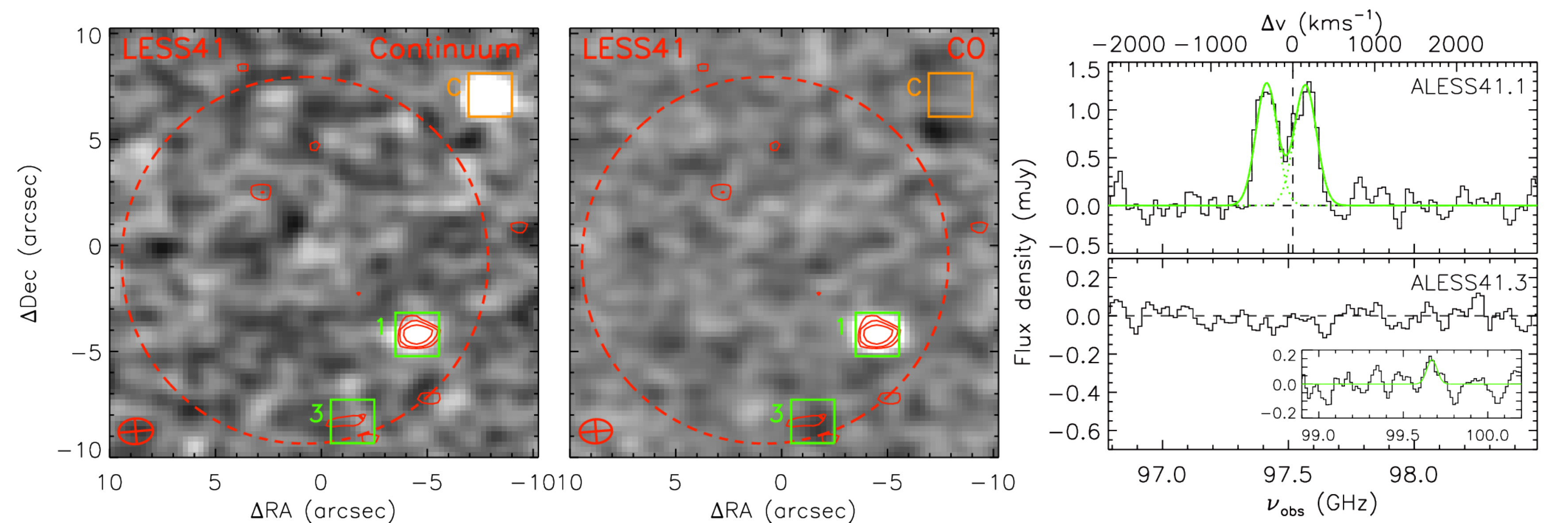}
\includegraphics[width=\textwidth]{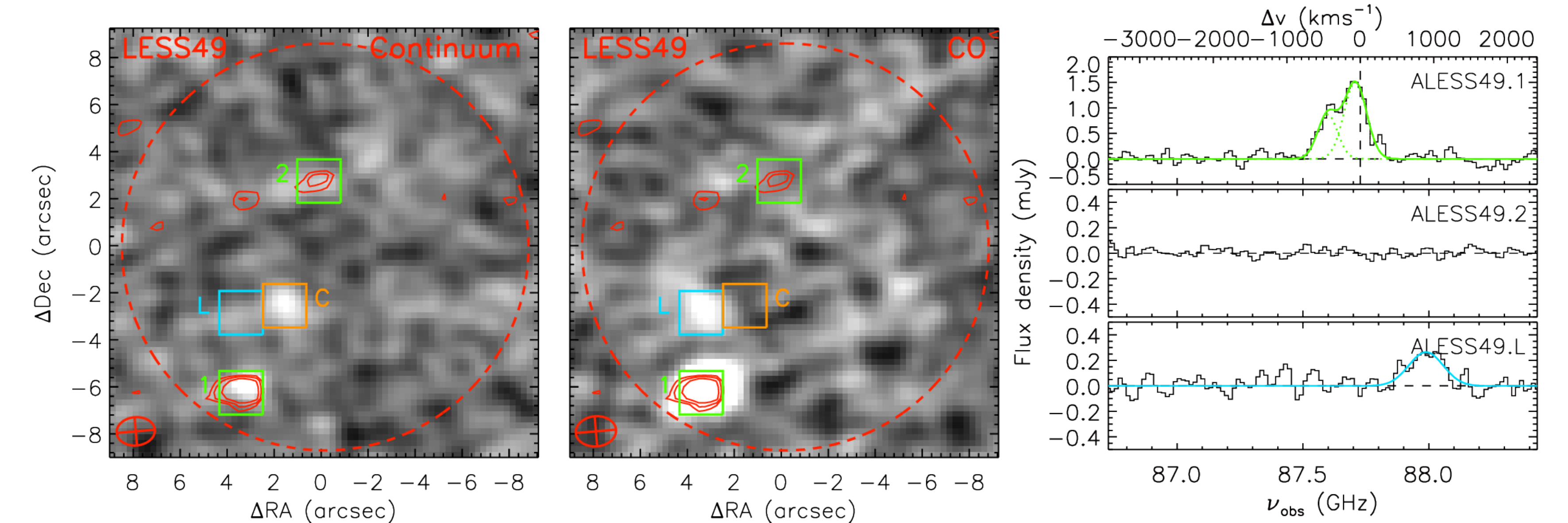}
\includegraphics[width=\textwidth]{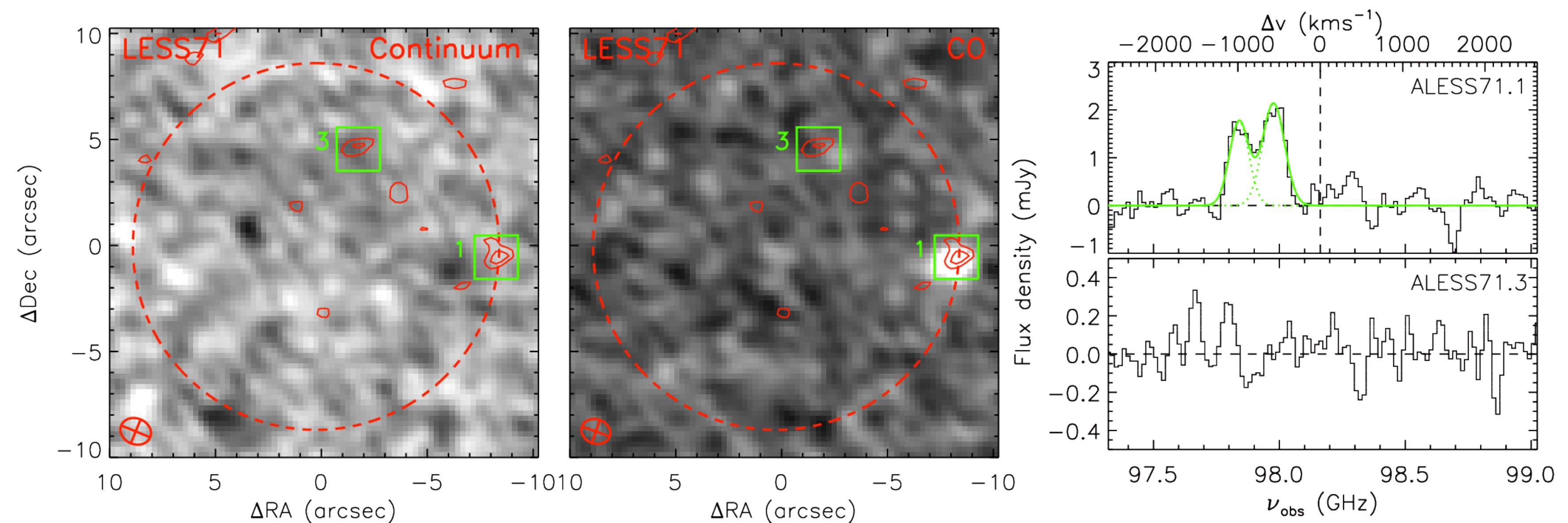}
\caption{An overview of the ALMA data in each field, demonstrating the quality of the data, the strength of the detections and the correspondence between the 870-\micron\ selected SMGs and sources in our 3.3-mm data. {\it Left column:} The naturally-weighted 3.3\,mm continuum data (greyscale), with 870-\micron\ continuum contours (at 2.5, 3.5, 5, 10, $20\sigma$) overlaid from \citet{Hodge13} or \citet{Simpson15b}. The dashed circle shows the 870\,\micron\ primary beam and the ellipse in the lower-left corner of each image represents the 3.3\,mm restored beam. At $\sim55\arcsec$ the 3.3\,mm primary beam is larger than the areas shown, but there are no significant sources outside of the areas shown. 870-\micron\ selected SMGs are highlighted and labelled with green boxes, 3.3-mm selected continuum sources (Section~\ref{sec:serendip_cont}) are marked in orange, and serendipitously-identified emission line sources (Section~\ref{sec:serendip_line}) are in cyan. Labels are the sub-IDs of the sources (Table~\ref{tab:sample}), so source `1' in the LESS\,41 panels is ALESS\,41.1 etc. 
{\it Centre column:} Collapsed, continuum-subtracted, naturally-weighted images covering the frequencies of the emission lines (i.e.\ moment-0 maps of the line emission).  Annotations and labels are the same as in the left-hand panel. 
{\it Right column:} Continuum-subtracted spectrum for each SMG and serendipitously-detected emission line source; labels in the top-right of each panel show the name of the source. Dashed vertical lines mark the expected frequency of CO emission based on spectroscopic optical/near-infrared redshifts, where they exist (Table~\ref{tab:sample}), and the top axis shows the velocity offset from this frequency. 
Best-fit single or double Gaussian profile are overlaid for sources with detected emission lines (Section~\ref{sec:findsmglines} and \ref{sec:serendip_line}). ALESS\,41.3 includes an inset showing the possible weak line in the other sideband at $\sim99.7$\,GHz (Section~\ref{sec:findsmglines}).
 }
\label{fig:overview}
\end{figure*}
\begin{figure*}
\includegraphics[width=\textwidth]{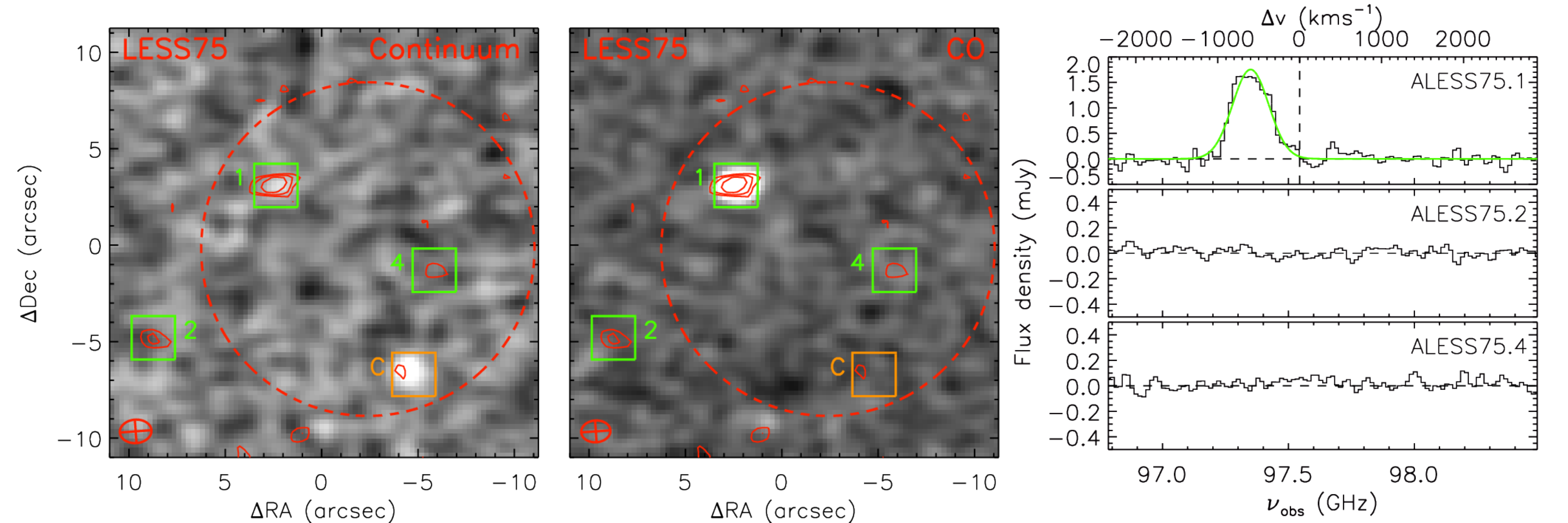}
\includegraphics[width=\textwidth]{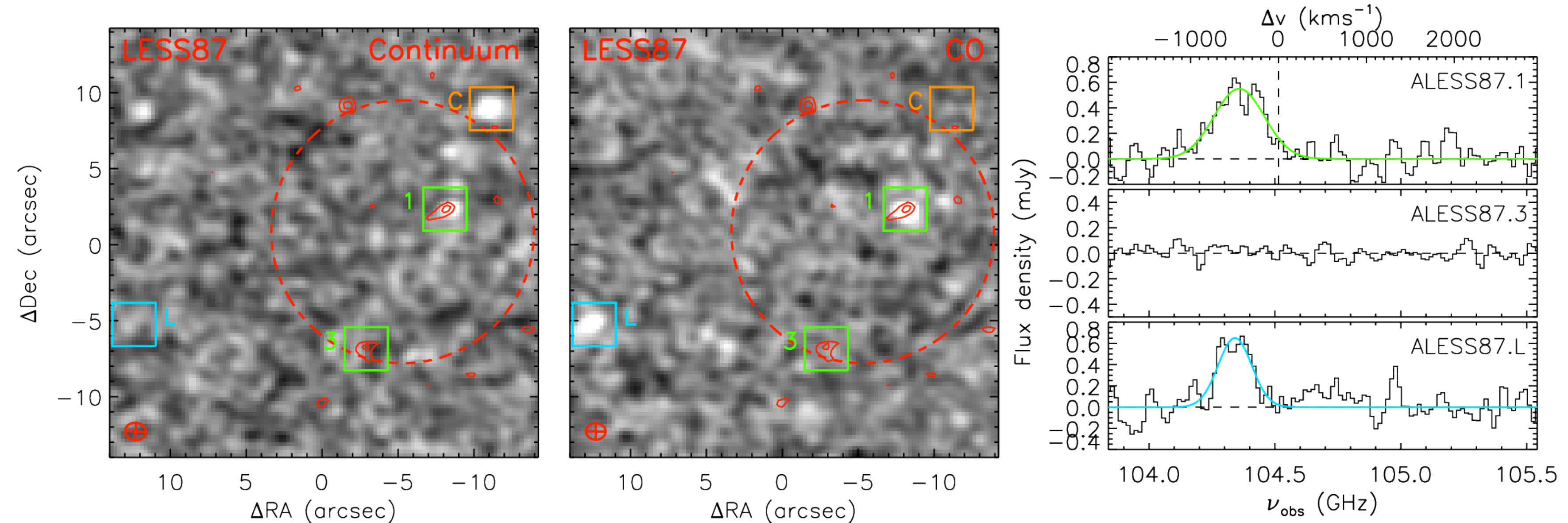}
\includegraphics[width=\textwidth]{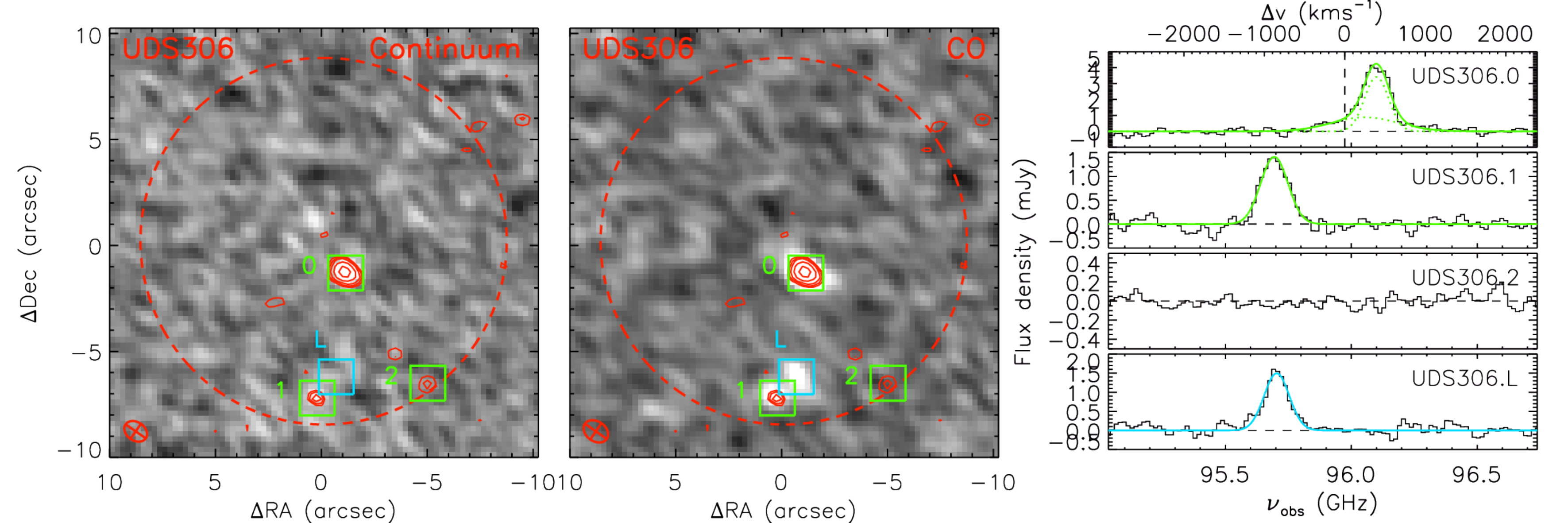}
\contcaption{}
\end{figure*}

\subsection{Identifying CO emission from the SMGs}
\label{sec:findsmglines}

To identify whether the 870-\micron\ selected SMGs exhibit any CO emission in our ALMA observations we begin by extracting the spectra in both sidebands from the dirty maps at the positions of the SMGs, average over the synthesised beam area and subtract any continuum. We then rebin these spectra to approximately 50, 100, and 200\,\kms\ velocity bins (corresponding to binning factors of 1, i.e.\ native resolution, $2\times$ and $4\times$) and search for $\ge2\sigma$ peaks in the rebinned spectra. A Gaussian profile is fit to each potential line and the integrated signal-to-noise ratio (SNR) measured from the rebinned spectra over twice the expected FWHM. For each of the three velocity binnings this procedure is repeated on the inverted spectra and the SNR at which the false detection rate (FDR) drops to zero is used as the threshold for identifying emission lines from the SMGs. The integrated line fluxes and widths are measured using the procedure described in Section~\ref{sec:linemeasure}.

Seven of the 870-\micron\ selected SMGs are detected as CO emitters with SNR at least twice the zero FDR threshold in all three velocity binnings employed (Table~\ref{tab:sample}; Figure~\ref{fig:overview}). The CO and optical/near-infrared redshifts agree within the expected uncertainties (to within 800\,\kms\ in all cases\footnote{Several of the optical/near-infrared redshifts are based on Lyman-$\alpha$ emission, which can be significantly offset from the systemic redshift \citep[e.g.][]{Erb14, Shibuya14}}) for the six SMGs with existing spectroscopic redshifts (Table~\ref{tab:sample}), confirming the reliability of multi-line optical/near-infrared spectroscopic redshifts for SMGs.

We also see a possible line at 99.7\,GHz at the position of ALESS\,41.3 (Figure~\ref{fig:overview}), which only exceeds the detection threshold when the fitting is performed on the unbinned data (i.e.\ at 50\,\kms\ resolution), and then has an SNR of 8.53, just above the false-detection threshold of 8.50. If real, and \cco\ emission, this line is offset by 6700\,\kms\ (2.2\,GHz) from the \cco\ line of ALESS\,41.1.
However, given that the significance of this line is at the limit of our sample we do not include it in our subsequent analyses.

\subsection{Serendipitously-detected emission line sources}
\label{sec:serendip_line}

Whilst the primary goal of our ALMA observations was to measure the redshifts and $^{12}$CO properties of the multiple components of the 870-\micron\ selected SMGs, we also search for additional line emitters in the ALMA cubes. 
We begin with the naturally-weighted ALMA cubes, rebinned to scales of $\sim0.6\arcsec~{\rm pixel}^{-1}$, such that the maps are Nyquist sampled. We mask the areas within 1\arcsec\ (i.e.\ approximately the ALMA beam) of the known SMGs and any serendipitously-detected continuum sources and then collapse both the sidebands of the cubes into slices 200\,\kms\ wide in steps of 100\,\kms. For each slice we search within the primary beam for pixels brighter than $5.5$ times the rms noise of the slice, a threshold chosen so that the number of false positives (based on performing the same search on the inverted cubes) is zero. We limit our search to the primary beam area; outside of the primary beam the false detection rate is substantially higher. We also perform a secondary search, with a detection limit of $5.0\sigma$, but only at the positions of 3.6\,\micron\ sources (from the  Spitzer IRAC/MUSYC Public Legacy in ECDF-S survey [SIMPLE; \citealt{Damen11}]). This secondary search has a false contamination rate of 15\% (calculated based on the statistics of the inverted cubes and the IRAC source density). 

We identify two line emitters in the unbiased $5.5\sigma$ search -- one in each of the LESS\,87 and UDS\,306 cubes, at 7.4 and $13.9\sigma$, respectively -- and one line emitter in the 3.6-\micron\ prior search, which is in the LESS\,49 field at the $5.0\sigma$ level.
We discuss these sources further in Sections~\ref{sec:assoclines} and \ref{sec:environ}.

\subsection{Line flux measurements}
\label{sec:linemeasure}

Once an emission line source is identified in the ALMA maps we extract and measure the properties of the emission line in the spectrum. This is undertaken using an iterative procedure on the naturally-weighted continuum-subtracted cubes as follows: we first extract the spectra in the beam area centred on either the 870-\micron\ position (for 870-\micron\ selected SMGs) or the peak of the brightest channel emission (for serendipitously-detected line sources). Based on these extractions we identify channels containing line emission and collapse the cubes over those channels to create a 2D map of the line. We then iterate to re-extract the spectra based on the areas containing emission in the collapsed map and re-identify line-containing channels to create a new map. Typically, only two iterations are required for the line-containing channels and areas to converge, and it is these areas over which we extract the final science spectra. These spectra are shown in Figure~\ref{fig:overview}. For 870-\micron\ selected SMGs without emission lines in our data (Section~\ref{sec:findsmglines}) the final science spectra are those extracted in the beam area at the 870-\micron\ position. 

The extracted spectra are fit with single or double Gaussian profiles using {\sc mpfit} in {\sc IDL} \citep{Markwardt09}. Uncertainties on the fit and derived parameters are determined from 1000 trials for each line, where random noise with the same $1\sigma$ rms as the spectra is added and the line refit. For 870-\micron\ selected sources without detected line emission we determine the integrated line flux limit at which an emission line with a  Gaussian profile and 500\,\kms\ linewidth would have been detected above the zero FDR threshold described in Section~\ref{sec:findsmglines}. The measured line fluxes, widths, and CO redshifts are listed in Table~\ref{tab:sample}.

\subsection{Serendipitously-detected continuum sources}
\label{sec:serendip_cont}

In addition to measuring the continuum fluxes from the known (i.e.\ 870-\micron\ selected) SMGs we search for additional 3.3-mm continuum sources in the maps, which are $\sim10\times$ larger by area than those obtained at 870\,\micron. Four 3.3-mm continuum emitters are identified that were not significantly detected in the original 870-\micron\ ALMA data -- one in each of the LESS\,41, LESS\,49, LESS\,75 and LESS\,87 fields. Using the same procedure described in Section~\ref{sec:findsmglines}, we determine that none of their spectra contain any detectable emission lines in the observed frequency range. Hereafter, these serendipitously-identified 3.3-mm continuum selected sources are named ALESS\,41.C, ALESS\,49.C, ALESS\,75.C and ALESS\,87.C, respectively (Table~\ref{tab:sample} and Figure~\ref{fig:overview}). 
The 3.3-mm flux densities for these sources are measured as described in Section~\ref{sec:almaobs} and presented in Table~\ref{tab:sample}. A discussion of the multiwavelength properties of these sources is presented in Appendix~\ref{sec:contdetail}. 

We next consider the surface density of the serendipitously-detected continuum sources to investigate whether there is an overdensity in the SMG fields.
Of the four 870-\micron\ faint, 3.3mm continuum sources detected, three are sufficiently bright to be detectable above $3.5\sigma$ in all six of the  pointings, i.e. we have observed three 3.3-mm sources at $\ge50~\mu$Jy in $\sim6~$arcmin$^2$, or four 3.3-mm sources at $\ge38~\mu$Jy in $\sim5~$arcmin$^2$. 
Few comparable blank-field 3\,mm continuum surveys have been published to date, and as such only the ALMA SPECtroscopic Survey in the Hubble Ultra-Deep Field \citep[ASPECS; ][]{Walter16, Aravena16a, Decarli16a, Decarli16b} provides a suitable comparison sample. 
ASPECS performed a single-pointing blank field survey in ALMA Band 3, covering roughly $1\,\rm{arcmin^2}$, across the whole 31\,GHz Band 3 bandwidth, with typical noise of $\sim0.15$\,mJy\,beam$^{-1}$ per 50--60\,\kms\ channel, corresponding to a typical continuum noise level of 3.8~$\mu {\rm Jy~beam}^{-1}$. ASPECS therefore covers approximately one sixth of the area of our study to similar depths as our deepest continuum data. ASPECS detected only one 3\,mm continuum source, which has a 3\,mm flux density of around $30\,\mu$Jy. We conclude that detecting four serendipitous continuum sources in our data is marginally in excess of the blank-field expectations. However, uncertainties in the number counts are significant and therefore we cannot be sure which, if any, of our serendipitously-detected continuum sources are associated with the 870-\micron\ selected SMGs. 
However, the detection of a 3-mm bright continuum sources within 2\arcsec\ of ALESS49.L is a priori unlikely and so these sources may be associated.

\section{Analysis and Discussion}
\label{sec:analysis}

\begin{figure}
\includegraphics[width=8.8cm]{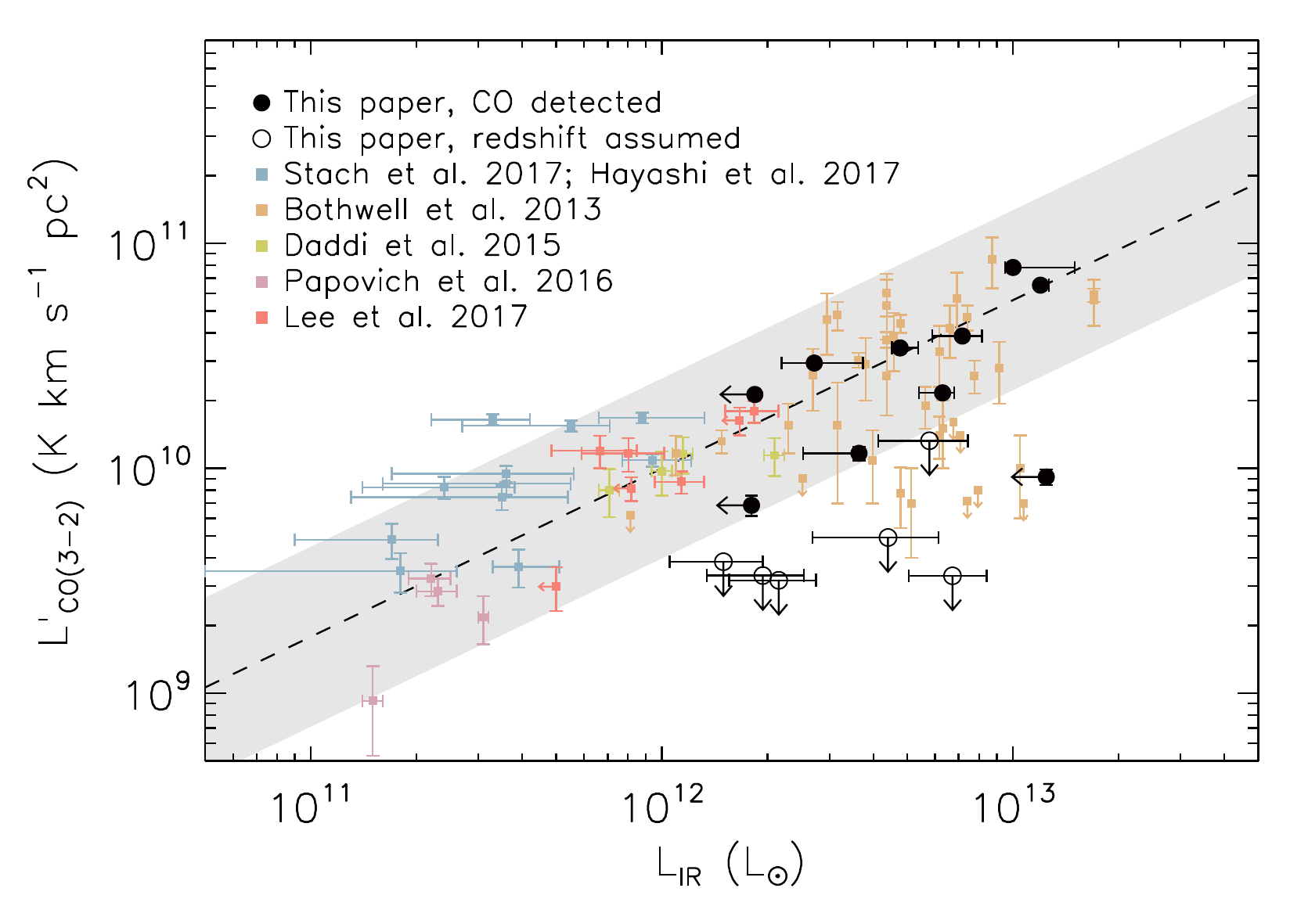}
\caption{Correlation between \cco\ and infrared luminosity for our 870-\micron\ selected SMGs and serendipitously-detected line emitters.
Our CO-detected 870-\micron\ selected SMGs have \cco\ luminosities consistent with expectations based on the trend with infrared luminosity from star-forming galaxies and previously-studied SMGs. 
To determine whether 870-\micron\ selected SMGs without detected CO emission could be physically associated with the other SMG(s) in each field we plot the limits for these CO-undetected SMGs, assuming that they are at the same redshift as the CO-detected companions. 
Also included for comparison are SMGs and other unlensed $z>1$ star-forming galaxies with \bco, \cco, or \dco\ measurements. The dashed line and grey shaded region show the linear fit and measured $1\sigma$ scatter of 0.4\,dex between $L^{\prime}_{\rm CO(3-2)}$ and $L_{\rm IR}$ for these galaxies, and represent the range expected for our sources. 
Most of our CO-undetected SMGs would have to have \cco\ luminosities significantly lower than this trend, and so be extremely gas poor, if they are at the same redshift as  the other SMG(s) in the field. We conclude that (with the possible exception of ALESS\,71.3, which has the highest limit on \cco\ luminosity) the CO-undetected SMGs are likely to lie at redshifts that place their CO emission outside of the ALMA bandwidth of our observations. 
Also note that even though our serendipitously-detected line emitters were not previously detected in the infrared continuum, they all have infrared luminosity limits that are consistent with the $L^{\prime}_{\rm CO(3-2)}$--$L_{\rm IR}$ relation.
}
\label{fig:lirlco}
\end{figure}

\begin{figure}
\includegraphics[width=8.8cm]{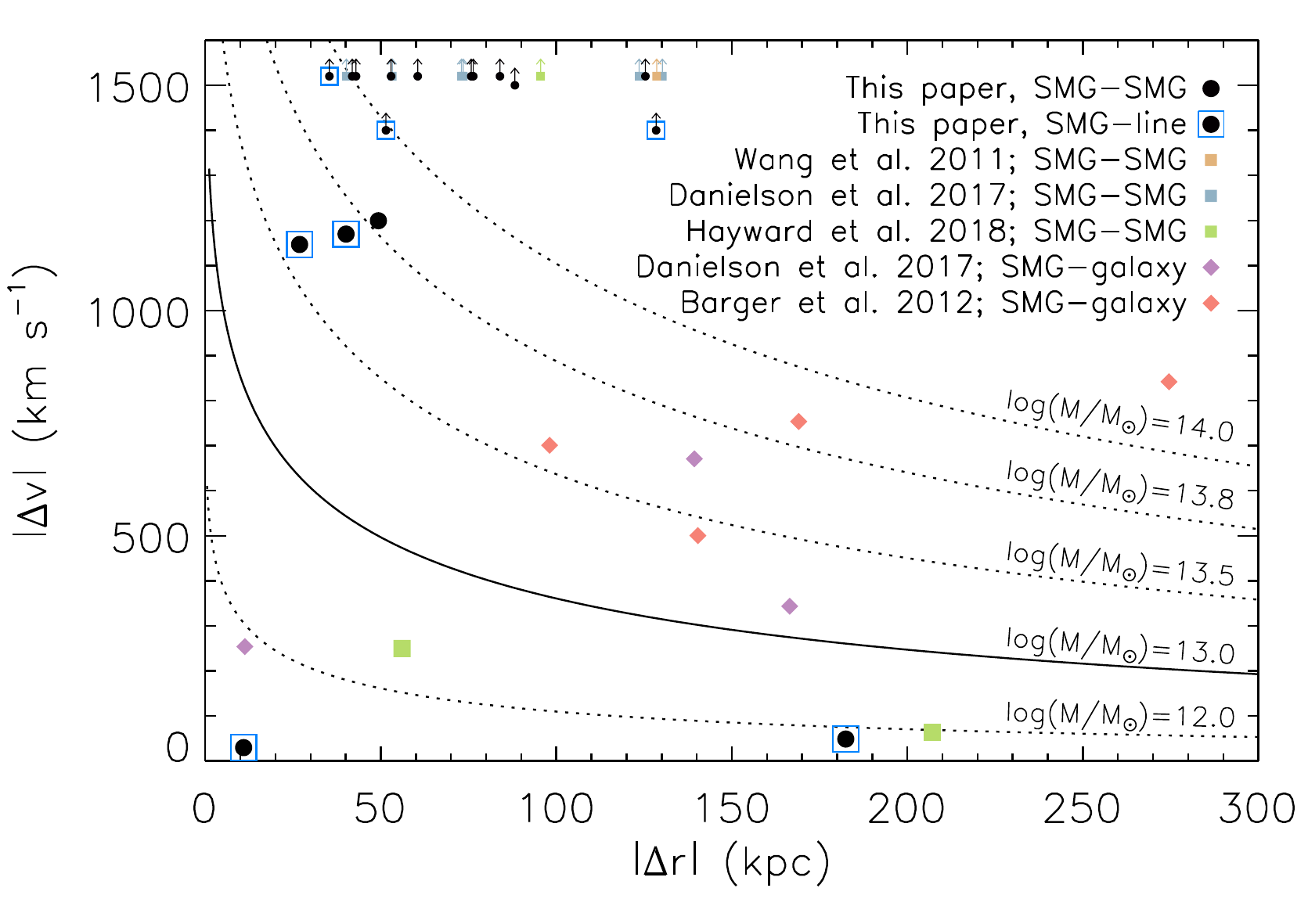}
\caption{Radial and velocity separations between the multiple 870-\micron\ selected SMGs (SMG--SMG) as well as serendipitously-detected line emitters (SMG--CO) in each ALMA field. The abscissa is truncated such that galaxy pairs with velocity separations (or lower limits) $>1500$\,\kms\ are shown as limits at 1500\,\kms. The curves show the expected profiles for NFW halos of different masses (see Section~\ref{sec:blends} for details). Clustering measurements suggest that 870-\micron\ selected SMGs reside in halos of mass $\sim10^{13}$\,\msun\ at $z\gtrsim2$, and simulations suggest that at $z\sim2.5$ virialised halos with mass $\gtrsim10^{14}$\,\msun\ are rare. Therefore, the lower limits on the velocity offsets between blended SMG--SMGs pairs suggest that they are unlikely to reside in the same bound environments. For comparison we also show SMG pairs and SMG--galaxy pairs from blank-field surveys that have interferometric positions and spectroscopic redshifts. The majority of the associated SMG--CO sources, SMG--SMG and SMG--galaxy pairs have velocity offsets higher than expected if they occupy the same virialised $\sim10^{13}$\,\msun\ halo, and may instead indicate that they trace different protocluster substructure.
Measurement uncertainties are not shown because they are typically smaller than the data points. 
}
\label{fig:dvdr}
\end{figure}

The correlation between \cco\ and infrared luminosity (Figure~\ref{fig:lirlco}), which is observed over several orders of magnitude in both local and high-redshift galaxies, reflects  the relationship between star-formation rate (traced by infrared luminosity) and gas mass (traced by \cco\ luminosity). This trend is similar to the observed correlation between \aco\ luminosity and infrared luminosity that  holds for galaxies across many orders of magnitude locally, and for those $z>0$ galaxies that have been observed in \aco\ \citep[e.g.][]{Schmidt59, Solomon97, Kennicutt98a, Ivison11, Daddi10a, Saintonge12, Huynh17}.

In Figure~\ref{fig:lirlco} we show the \cco\ luminosity ($L^{\prime}_{\rm CO(3-2)}$) and infrared luminosity ($L_{\rm IR}$) derived from the 250--870\,\micron\ photometry \citep{Swinbank14} for 870-\micron\ selected SMGs with detected $^{12}$CO. Included for comparison are star-forming galaxies with \bco, \cco, or \dco\ measurements \citep{Bothwell13, Daddi15, Papovich16, Lee17, Stach17, Hayashi17}. Where \cco\ observations are unavailable, \bco\ and \dco\ luminosities are converted to $L^{\prime}_{\rm CO(3-2)}$ using the brightness temperature ratios for the SMGs compiled by \citet{CarilliWalter13}. Only \bco, \cco, and \dco\ data are included in order to minimise uncertainties from the extrapolation of the CO spectral line energy distribution (SLED).

Figure~\ref{fig:lirlco} shows that the \cco\ luminosities of the seven 870-\micron\ selected SMGs from which we detect $^{12}$CO emission are all consistent with the previously-observed $L^{\prime}_{\rm CO(3-2)}$--$L_{\rm IR}$ correlation. This suggests that the gas properties of these galaxies are similar to previously-studied SMG populations. 

We next use our $^{12}$CO data to investigate the association between the multiple components in blended single-dish submillimetre sources, the new serendipitously-detected line emitters, and the environments  of SMGs.

\subsection{Are blended SMGs physically associated?}
\label{sec:blends}

Here, we use our ALMA data to determine whether the multiple 870-\micron\ selected SMGs that are blended in single-dish survey data are physically associated or are instead chance, line-of-sight projections. In only one of our six target fields was $^{12}$CO detected from more than one 870-\micron\ selected SMG  (UDS\,306). For the six other SMGs without existing spectroscopic redshifts $^{12}$CO is not detected in our observations (Section~\ref{sec:findsmglines}), and therefore we cannot confirm their redshifts. Instead, we use the depth of our ALMA data and the $\sim5000$\,\kms\ coverage of the ALMA sidebands to determine whether these SMGs would have been detected if they were physically associated with the SMGs that have optical/near-infrared spectroscopic redshifts. 
There are two aspects of this test, which we discuss next: 1) examining whether the observations are deep enough to detect a $^{12}$CO line; and 2) determining whether the bandwidth was wide enough to include CO emission from a physically associated companion.

Figure~\ref{fig:lirlco} includes limits on the CO-undetected SMGs in our sample, assuming that they have a redshift that would place CO within $\sim2500$\,\kms\ of the spectroscopically-confirmed SMG(s). 
Based on the trend and scatter between $L^{\prime}_{\rm CO(3-2)}$ and $L_{\rm IR}$ for star-forming galaxies (including SMGs) it is clear that {\it if} the secondary SMGs are at the same redshift as the primary SMG in each field then they have CO-luminosities significantly lower than predicted based on their infrared luminosities. Indeed, in this scenario, five of our seven secondary SMGs would have CO(3-2) luminosities (and thus gas masses) approximately a factor of two lower than {\it any} known SMG with comparable infrared luminosity, and one other (ALESS71.3) would be fainter than about 70\% of SMGs. 
Since it is highly unlikely that our sample of seven SMGs includes the five faintest CO(3-2) SMGs, and another in the faintest 30\%, we conclude that (with the possible exception of ALESS71.3, which has the least stringent limit) these CO non-detections are most likely not a result of the observations being too shallow to detect any emission, but instead that their CO emission lies outside of the ALMA bandwidth. Thus, these SMGs have redshifts that are different to the spatial companions with which they are blended in single-dish continuum data. 

Figure~\ref{fig:dvdr} shows the projected velocity and spatial separation between SMGs and nearby companions, and is used to address whether the ALMA observations covered a sufficiently wide bandwidth to include CO emission from physically-associated companions. Curves showing the expected velocity separation of pairs of test masses in Navarro-Frenk-White (NFW; \citealt{Navarro97}) halos are included, calculated using the formalism of \citet{Lokas01} and assuming a concentration parameter of 3.5, as expected for halos of mass  $10^{12-14}$\,\msun\ at $z=2$--3 \citep{Dutton14}. There are 10 SMG--SMG pairs in our six ALMA-targeted fields  (note that the UDS\,306 field contains three SMGs, and therefore three SMG--SMG points are displayed: UDS\,306.0$\rightarrow$UDS\,306.1;  UDS\,306.0$\rightarrow$UDS\,306.2; and UDS\,306.1$\rightarrow$UDS\,306.2).
As discussed above, in SMGs where CO is undetected we can exclude redshift solutions that would place CO emission within the ALMA bandpass. Therefore, SMG pairs that include one galaxy with an unknown redshift are plotted as lower limits on their velocity separation. 

The only SMG--SMG pair with CO detected in both components (UDS\,306.0 and UDS\,306.1) has a velocity separation of 1200\,\kms\ across 50\,kpc. If these galaxies are within the same virialised halo and trace the gravitational potential of the halo, then it has a likely mass $\sim10^{13.8}$\,\msun. All other SMG--SMG pairs have velocity separations $>1500$\,\kms\ (and most $>2000$\,\kms; Figure~\ref{fig:overview}). Therefore, if these single-dish blended SMGs reside in the same virialised halos and have relative motions dominated by the halo gravity, then the halos have masses $\gg10^{14}$\,\msun. At $z\sim3$ such massive halos are expected to be very rare, $\sim10^{-9}\,{\rm Mpc}^{-3}$ \citep[i.e.\ only $\sim10^{-4}$ halos of mass $>10^{14}$\,\msun\ are expected in the whole ECDFS field at these redshifts; e.g.][]{Murray13}. 

We therefore conclude that the SMGs in our study without detected CO are unlikely to be closely physically associated with those with confirmed redshifts (with the possible exception of ALESS\,71.3, for which low-luminosity CO is possible, as discussed above), and that ongoing interactions between the blended SMG--SMG pairs are unlikely to have triggered the high star-formation rates in these galaxies (Section~\ref{sec:environ}). We conclude that at most one [or two; allowing for low luminosity CO in ALESS\,71.3] of our six ($17\%$ [33\%]) blended SMG fields contains a bound SMG pair. 
There are a further five blended 870-\micron\ sources in single-dish blank-field surveys with spectroscopic redshifts for both constituent SMGs (\citealt{Wang11, Danielson17, Hayward18}\footnote{We take the conservative approach and only consider SMGs selected from single-dish surveys at 850--870\micron\ with robust  spectroscopic redshifts for multiple components. We require spectroscopic redshifts for both components because the currently available literature datasets are based on optical/near-infrared spectroscopy, where lines may not be detected from two galaxies at the same redshift due to dust absorption or intrinsically faint emission. This  selection results in the addition of only three of the \citet{Hayward18} sample. Note that the spectroscopic redshift of component `a' of GOODS\,850-15 listed by \citet{Hayward18} is actually that of component `b'  \citep{Chapman05}, the other putative member of the pair. The redshift appears to have been erroneously applied to component `a' by \citet{Barger14}. No spectroscopic redshift exists for component `a' of GOODS\,850-15.}), which are also shown in Figure~\ref{fig:dvdr}. Including these sources in the statistics reveals that only three [four] in 11 ($27\%$ [36\%]) blended SMG fields contains closely physically associated SMGs ($\lesssim300$\,kpc and $\lesssim1500$\,\kms). 
This result is in tension with the high number of blended submillimetre sources in blank field surveys (Section~\ref{sec:intro}; \citealt{Simpson15b}), which we discuss in Section~\ref{sec:environ}, in the context of the wide-field environment of SMGs.

\subsection{What are the serendipitously-detected line emitters?}
\label{sec:assoclines}

We next investigate the serendipitously-detected line emitters that were detected in our ALMA spectroscopy. 
Due to the frequency range of our observations (approximately 7.5\,GHz across the two sidebands) we consider it most likely that the serendipitously-detected emission lines (Section~\ref{sec:serendip_line}) are the same lines as targeted for the SMG in each field, i.e.\ \cco\ for all the fields with serendipitously-detected lines. This is because lower $J$ CO lines probe lower redshifts and thus significantly smaller volumes (roughly 11\,Mpc$^3$  per field at $z\sim0.3$ for \aco, and 250\,Mpc$^3$ per field at $z\sim1.4$ for \bco, compared with 440\,Mpc$^3$ at $z\sim2.5$ for \cco). \eco\ and higher $J$ lines probe $z>4.9$ and are therefore unlikely due to the low volume density of luminous galaxies at those redshifts. Similarly, detectable atomic transitions require high redshift and high luminosity galaxies. Thus, \cco\ at the targeted redshift (there are no serendipitously-detected emission-line sources in the LESS\,71 field, which we targeted in \dco), or \dco\ at $z\sim3.8$ are the most likely candidates based on the volume and luminosity. Furthermore, if the emission-line sources are unassociated with the targeted SMGs then we expect an approximately equal number of serendipitous line sources in both sidebands. Instead all three of the serendipitous line sources are observed in the same sideband as the CO from the targeted 870-\micron\ selected SMGs (Section~\ref{sec:serendip_line}), making the likely emission \cco\ associated with the target SMG (12\% random chance). We note that the optical/near-infrared photometry of the serendipitously-detected emission-line sources (Almaini et al.\ in prep.) is consistent with the redshifts inferred from this conclusion.

We also use ALMA blank-field emission-line studies to estimate the number of submillimetre-faint CO line emitters that are expected in a survey of the same volume as our observations. As discussed in Section~\ref{sec:serendip_cont}, ASPECS \citep{Walter16, Aravena16b, Decarli16a, Decarli16b} observed approximately 1.3~times the combined spatial and spectral area of our study in ALMA Band 3 (one sixth of the spatial area, but eight times the frequency coverage), although ASPECS is typically 3~times deeper. 
ASPECS detected ten CO emitting sources in their blank-field survey (with $\le4$ expected to be spurious), of which half would be detectable at $\ge5\sigma$ in our five deepest fields (where all three of our serendipitously-detected line sources are located), and two in all of our observations. Based on ASPECS we therefore expect to detect 0.9--2.3 sources in both sidebands of all our fields by chance (assuming a 40\% false-positive rate in ASPECS), i.e.\ 0--1 in one of the sidebands. Although there are large statistical uncertainties in this calculation, the comparison further supports our conclusion that a majority of the serendipitously-detected lines are likely to be \cco\ emission from galaxies associated with, and at similar redshifts to, the target SMGs. 

The conclusion that the serendipitously-detected line emitters are associated with the spectroscopically-confirmed 870-\micron\ selected SMGs is also supported by Figure~\ref{fig:lirlco}, which shows the correlation between infrared luminosity and \cco, and includes the serendipitously-detected line emitters, assuming that their emission is from \cco\ at a similar redshift to the targeted SMGs.
Limits on the infrared luminosities of the line emitters are determined by assuming the same $L_{\rm IR}/S_{870}$ ratios for the  870-\micron\ selected SMGs in each field, and extrapolating to the 870\micron\ detection limit. For sources within the ALMA 870-\micron\  primary beam the ALMA detection limits is used, and for ALESS\,87.L, which is $\sim10\arcsec$ outside of the ALMA primary beam, we employ the LABOCA detection limit. Thus, we implicitly assume that the SMGs and the serendipitously-detected line emitters have similar dust SEDs. 
It is clear from Figure~\ref{fig:lirlco} that if the line emission is indeed from \cco\ then in all cases the infrared luminosity limits are consistent with expectations based on the observed $L^{\prime}_{\rm CO(3-2)}$--$L_{\rm IR}$ trend and observed scatter.

\subsection{SMG environments and star-formation triggers}
\label{sec:environ}

Now that all the sources in our SMG fields have been investigated, we next use the spatial and spectral separation of pairs of galaxies to probe the large scale-environments of SMGs.
In addition to SMG--SMG pairs (Section~\ref{sec:blends}), in Figure~\ref{fig:dvdr} we also show the separations between SMGs and the serendipitously-detected emission-line sources (four pairs; hereafter SMG--CO pairs), and $z\sim2$--3 SMGs and other galaxies (seven pairs; hereafter SMG--galaxy pairs). We identify SMG--galaxy pairs  by cross-matching blank-field spectroscopic surveys of SMGs with nearby spectroscopically-identified field galaxies \citep{Barger08, Barger12, Danielson17}. The separations of these pairs can be compared with the expectations for galaxies orbiting in NFW halos of different masses (calculated as described in Section~\ref{sec:blends}). 

At $z=2.5$ SMGs typically have stellar masses of around $10^{11}$\,\msun\ \citep[e.g.][]{Wardlow11, Simpson14} and dynamical masses of a few times $10^{11}$\,\msun\ \citep[e.g.][]{Greve05, Swinbank06, Tacconi08, Bothwell13, Huynh17}. 
\citet{Patton13} found enhancement in the average star-formation rate of pairs of low-redshift galaxies compared with a matched control sample, for galaxy pairs with radial separations $<150$\,kpc, although the strongest enhancements are in the closest pairs (a factor of three, on average, for pair separations $\sim10$\,kpc). 
Simulations have shown that, whilst not all mergers induce starbursts, interaction-induced star-formation is strongest in pairs close to final coalescence, that galaxies with high velocity offsets ($\sim500$\,\kms) but small radial separations (tens of kpc) typically have the strongest induced starbursts, although there can be some enhancement to star-formation rates in the earlier stages of an interaction \citep[e.g.][]{DiMatteo08, Narayanan10, Sparre16}. 

One of our SMG--CO pairs (UDS\,306.1--UDS\,306.L), one SMG--SMG pair (from \citealt{Hayward18}), and one SMG--galaxy pair (from \citealt{Danielson17}) have radial separations $\ll150$\,kpc and velocity separations similar or less than typical for two galaxies in a halo of mass $\sim10^{12}$\,\msun. The star-formation in these three SMGs may therefore have been triggered by their interactions, but the spatial and spectral separations indicate that it is unlikely that the other 11 SMGs ($79\pm12\%$) with detected companions are triggered by interactions with the associated galaxies. 
The spatial resolutions of our data are $\sim1\arcsec$ (7.8\,kpc at $z=3$) and we trace galaxies with \cco\ luminosities of $\sim2\times10^{9}\,{\rm K~km~s^{-1}~pc^2}$ ($M_{\rm gas}=3\times10^9$\,\msun, assuming $L^{\prime}_{\rm CO(3-2)} / L^{\prime}_{\rm CO(1-0)}=0.66$ and $\alpha_{\rm CO}=1$\,\msun(K~\kms~pc$^2)^{-1}$). The spatial resolution of the other spectroscopic surveys considered here \citep{Wang11, Barger12, Danielson17, Hayward18} are similarly limited to galaxy pairs with $\gtrsim1\arcsec$ separations and are therefore similarly insensitive to late-stage mergers (where the components have already coalesced to sub-arcsec separations), minor-mergers, or the interactions with a gas-poor companion. Thus, interaction-triggering is possible in more than the three SMGs with both spectrally- and spatially-close companion galaxies.

Clustering studies have shown that $z=2$--3 SMGs typically reside in halos of mass $\sim10^{13}$\,\msun\ \citep{Hickox12, Wilkinson17}, although statistical uncertainties are significant (typically 0.3--0.5\,dex), and simulations suggest that there may be a systematic overestimation of a factor of $\sim2$--3, due to the effect of the blending of multiple galaxies in the large single-dish submillimetre survey beams \citep{Cowley17}.
Therefore, by treating all 14 spectroscopically-confirmed SMG pairs (three SMG--SMG, four SMG--CO, and seven SMG--galaxy) as an ensemble, we would expect them to scatter around or below the $M_{\rm halo}\sim10^{13}$\,\msun\ track on Figure~\ref{fig:dvdr}, if they reside in virialised environments\footnote{Note that if the halo masses of SMGs are systematically overestimated due to blending in the large submillimetre survey beams \citep{Cowley17}, and they reside in virialised environments, then we would expect them to scatter around a track between the $10^{12}$ and $10^{13}$\,\msun\ curves on Figure~\ref{fig:dvdr}. However, most of the SMGs are significantly above the $10^{13}$\,\msun\ track and therefore our conclusions are not affected if this "blending bias" in SMG halo masses is significant}. However, it is clear that the majority of the sample have higher velocity offsets and instead scatter around the velocity profile expected for a virialised $5\times10^{13}$\,\msun\ halo. Although this is somewhat larger than expected for the clustering-derived halo masses there are significant uncertainties in both analyses and therefore it is possible that SMG pairs reside in such environments, although we also note that virialised  $5\times10^{13}$\,\msun\ halos are rare at $z=2$--3 \citep[e.g.][]{Murray13}.

Halos of mass $10^{13}$\,\msun\ at $z=2.5$ (i.e.\ typical of SMGs) likely evolve into $10^{13.5-14}$\,\msun\ halos at $z=0$ \citep{Fakhouri10} -- the scales of local clusters and high-mass groups. Simulations have shown that the $z\sim2$--3 progenitors of such structures  are usually extended and filamentary in nature (covering tens of comoving Mpc)
\citep[e.g.][]{Chiang13, Muldrew15, Overzier16}. 
In Section~\ref{sec:blends} we showed that most SMG--SMG pairs do not appear to reside in the same virialised halo, but our frequency coverage is insufficient to determine whether the companion SMGs are members of the same larger-scale structure. Considering our data in conjunction with the determination that the fraction of blended SMGs is a factor of $\sim80$ higher than expected from chance line-of-sight alignments \citep{Simpson15b}, we hypothesise that the SMG pair components that we observe here may be tracing different filaments within the large-scale structure of the wider SMG environment. 

We also consider the broader spatial environment of the SMGs. The targets of our ALMA programme were selected on the basis of there being two or more SMGs within the $\sim15\arcsec$ single-dish beam (Section~\ref{sec:almaobs}), and the 3\,mm ALMA field of view is only $\sim60\arcsec$. Therefore, our targeted observations are more sensitive to line-of-sight structure than that in the plane of the sky. We can probe more spatially-extended structure by using the  original 870-\micron\ catalogue and existing photometric and spectroscopic redshifts (\citealt{Simpson14, Danielson17}; Almaini et al.\ in prep.) to identify any likely larger scale associations of SMGs. 
 ALESS\,87.1 is offset from ALESS\,122.1 by 39\arcsec\ (325\,kpc) and 615\,\kms\ in velocity, and ALESS\,49.1 is separated by $\sim100\arcsec$ (790\,kpc) and $\sim4100$\kms\ from ALESS\,107.1, suggesting that they may also be associated. This analysis indicates that the large-scale structures in which SMGs reside can be spatially as well as spectroscopically extended, and that the initial identification of more line-of-sight structure than that in the plane of the sky is a selection effect from our survey.

\section{Conclusions}
\label{sec:conc}

We presented the results from observations of single-dish selected submillimetre sources that are each comprised of two or more individual SMGs that are blended together in the large single-dish beam. Our main results and conclusions are as follows:  

\begin{enumerate}

\item We performed an ALMA CO survey of six 870-\micron\ selected single-dish submillimetre sources, which are comprised of the blends of  14 individual SMGs. We detected the targeted CO emission in all SMGs with previously known redshifts, but from only one of the seven companion SMGs. We also identified new, serendipitously-detected line-emitting sources in three of the six SMG fields, and new 3.3-mm continuum sources in four of the fields. Most of these serendipitously-detected sources are likely to be associated with the targeted SMGs. 
We also observed eleven ALMA-identified blended SMGs with XSHOOTER for optical/near-infrared spectroscopy, but those data did not yield redshifts for any of the targeted SMGs.

\item Our ALMA data are deep enough, and the bandwidths broad enough, to have detected CO from the companion SMGs if they were at the same redshift as the spectroscopically-identified SMGs, given their far-infrared luminosities. Therefore, in 5/6 cases (83\%) the multiple SMG components of blended single-dish submillimetre sources are not closely physically associated. We combined this result with another five blended submillimetre sources from the literature and concluded that $36(\pm18)\%$ of blended submillimetre  sources are comprised of multiple component SMGs that are closely physically associated ($\lesssim300$\,kpc and $\lesssim1500$\,\kms). In $64(\pm18)\%$ of cases the blended SMGs are line-of-sight alignments or residing in different regions of the same large-scale structure.

\item The new, serendipitously-detected emission-line sources are most likely \cco\ from companion galaxies at similar redshifts to the targeted SMGs. Similarly to associated SMG--galaxy pairs from the literature, these SMG-companions are likely tracing the same large scale structure as the SMGs. 

\item Only three of the 14 of SMGs with spectroscopically-confirmed companions  ($21\pm12\%$) have small enough spatial and spectral separations ($\sim8$--150\,kpc and $\lesssim300$\,\kms) that their star-formation may have been triggered by an interaction with the companion. Additional studies are needed to expressly investigate this possibility and to examine whether post-coalescent mergers and interactions with low-mass or gas-poor galaxies may account for the starbursts in the remaining $\sim80\%$ of SMGs.

\end{enumerate}

\section*{Acknowledgements}
We thank Ryley Hill for informative discussions. JLW acknowledges support from a STFC Ernest Rutherford Fellowship (ST/P004784/1) and from a European Union COFUND/Durham Junior Research Fellowship under EU grant agreement number 609412. IRS, EAC and BG acknowledge support from the ERC Advanced Grant DUSTYGAL (\#321334), and IRS also recognises support from a Royal Society/Wolfsom Merit Award. JLW, IRS, AMS, EAC and BG also acknowledge support from STFC (ST/P000541/1). HD acknowledges financial support from the Spanish Ministry of Economy and Competitiveness (MINECO) under the 2014 Ram\'on y Cajal program MINECO RYC-2014-15686, and JH acknowledges support of the VIDI research programme with project number 639.042.611, which is (partly) financed by the Netherlands Organisation for Scientific Research (NWO).
This paper makes use of the following ALMA data: ADS/JAO.ALMA\#2016.1.00754.S. ALMA is a partnership of ESO (representing its member states), NSF (USA) and NINS (Japan), together with NRC (Canada), NSC and ASIAA (Taiwan), and KASI (Republic of Korea), in cooperation with the Republic of Chile. The Joint ALMA Observatory is operated by ESO, AUI/NRAO and NAOJ.
Based in part on observations collected at the European Organisation for Astronomical Research in the Southern Hemisphere under ESO programme 094.A-0811(A).
The raw and pipeline-processed data used in this paper are available from the ALMA Science Archive, and the raw XSHOOTER data are available from the ESO Science Archive Facility.




\bibliographystyle{mnras}
\bibliography{/Users/jwardlow/papers/bibtex.bib} 


\appendix

\section{Multiwavelength properties of the serendipitously-detected 3.3\,mm continuum sources}
\label{sec:contdetail}

In this Appendix we provide detailed discussion of the multi-wavelength properties of the 3.3-mm selected continuum sources that were serendipitously-detected by our ALMA data (Section~\ref{sec:serendip_cont}), 

We begin by considering the 870\,\micron\ properties of these sources, noting that the relative depths of our new 3.3\,mm data and the existing 870\,\micron\ observations is such that high-redshift ($z\gtrsim3.5$) and/or cool (dust temperatures $\lesssim25$\,K) thermal sources with power law emissivity index, $\beta\sim1.5$ and 3.3\,mm fluxes, $S_{\rm 3mm}\lesssim60\mu$Jy, may not be detected in the 870\,\micron\ data. i.e.\ the mere detection of a new continuum source at 3.3\,mm is insufficient to infer that the source is non-thermal in nature -- instead the source brightnesses and available multiwavelength data must also be considered.  

ALESS\,41.C and ALESS\,87.C are both outside of the primary beam of the original 870-\micron\ ALMA observations and ALESS\,41.C  corresponds to a $\sim2\sigma$ peak in that map, suggesting that there may be some flux just below the threshold of the 870-\micron\ data. ALESS\,75.C and ALESS\,49.C are both inside the 870-\micron\ primary beam and ALESS\,75.C is coincident with a 2--$3\sigma$ region in that map and so might have been detected in marginally deeper data. There is no indication of any emission in the 870-\micron\ data at the positions of ALESS\,49.C or ALESS\,87.C. In all of our fields the 870-\micron\ flux from the newly-identified 3.3-mm selected sources (Table~\ref{tab:sample}), when summed with the flux from the 870-\micron\ selected SMGs does not significantly exceed the original single-dish LABOCA fluxes. However, the uncertainties are substantial and as such, in all cases, the 870-\micron\ selected SMGs alone are also sufficient to account for all of the LABOCA-measured flux, within the errors.

ALESS\,41.C has a 3.3\,mm/870\,\micron\ flux ratio approximately three times higher than typical modified blackbody spectral energy distributions (SEDs) and corresponds to the known radio emitter SWIRE~J033109.80$-$275225.3, with $S_{\rm 1.4GHz}=0.19\pm0.04$\,mJy \citep{Norris06}. There is no archival redshift available, and it has an infrared/radio SED consistent with non-thermal emission from a flat-spectrum radio quasar (FSRQ). ALESS\,41.C lies outside of the coverage of both the deep CDFS and shallower, wider ECDFS {\it Chandra} X-ray data \citep{Lehmer05, Virani06, Luo08, Luo17}

ALESS\,49.C is not detected in the near-infrared ($J$ and $K$ to $5\sigma$ depth of 25.3 AB mag from the Taiwan ECDFS Near-Infrared Survey [TENIS; \citealt{Hsieh12}], or IRAC from SIMPLE; \citealt{Damen11}) or the 250\,ks \chandra\ X-ray data. It is outside of the Multiwavelength Survey by Yale-Chile (MUSYC; \citealt{Taylor09}) area, and not significantly detected at above 20\,$\mu$Jy 1.4~GHz  \citep{Miller08, Biggs11}. The limit on the 3\,mm/870\,\micron\ flux density ratio is consistent with thermal emission at the the redshift of the SMG if the dust temperature is $\lesssim25$\,K, or warmer dust in a higher redshift galaxy. ALESS\,49.C is $<2\arcsec$ away from ALESS\,49.L (Section~\ref{sec:serendip_line}), indicating that it may be associated with the SMG, although the currently available data are ambiguous.

ALESS\,75.C is undetected at 1.4~GHz (to 20\,$\mu$Jy) and in the  250\,ks \chandra\ X-ray observation, but is detected by MUSYC and SIMPLE/IRAC, with a catalogued photometric redshift of $4.00^{+0.07}_{-0.08}$ \citep{Taylor09, Cardamone10}. The IR/radio SED, including the 3.3\,mm photometry and the 870\,\micron\ and 1.4\,GHz non-detections, is consistent with a cool ($\lesssim25$\,K) dusty galaxy at this redshift. 

Finally, ALESS\,87.C is detected with both IRAC/SIMPLE and MIPS/FIDEL \citep{Magnelli09} data, although it is outside the area of much of the ECDFS multiwavelength data, including the X-ray, radio and MUSYC. Therefore, we are unable to constrain the nature of ALESS\,87.C with the current data, although we note that the 3\,mm/870\,\micron\ ratio requires cool dust $\lesssim30$\,K and high redshift ($\gtrsim4$) if the emission is thermal.

\bsp	
\label{lastpage}
\end{document}